\documentstyle[12pt,epsfig]{article}

\begin{document}  
\vspace*{-2cm}  
\renewcommand{\thefootnote}{\fnsymbol{footnote}}  
\begin{flushright}  
hep-ph/0001295\\
PSI-PR-00-03\\  
January 2000\\  
\end{flushright}  
\vskip 65pt  
\begin{center}  
{\Large \bf The static QCD potential in coordinate space with quark masses
through two loops}\\
\vspace{1.2cm} 
{\bf  
Michael Melles\footnote{Michael.Melles@psi.ch}  
}\\  
\vspace{10pt}  
Paul Scherrer Institute,  
CH-5232 Villigen, PSI\\  
  
\vspace{70pt}  
\begin{abstract}
The potential between infinitely heavy quarks in a color singlet state is
of fundamental importance in QCD. While the confining long distance part
is inherently non-perturbative, the short-distance (Coulomb-like) regime
is accessible through perturbative means. In this paper we present new
results of the short distance potential in coordinate space with quark
masses through two loops. The results are given in explicit form based
on reconstructed  
solutions in momentum space in the Euclidean
regime. Thus, a comparison with lattice results in the overlap region
between the perturbative and non-perturbative regime is now possible
with massive quarks.
We also discuss the definition of the strong coupling based on the
force between the static sources.
\end{abstract}
\end{center}  
\vskip12pt

\setcounter{footnote}{0}  
\renewcommand{\thefootnote}{\arabic{footnote}}  
  
\vfill  
\clearpage  
\setcounter{page}{1}  
\pagestyle{plain} 
 
\section{Introduction} 

The potential between two (infinitely) 
heavy color charges in a singlet state has been
subject
to theoretical investigations for more than twenty years \cite{sus,fis,adm,
fei,bil}. 
In the non-perturbative
regime it is expected to play a key role in the understanding of quark 
confinement and it is a major ingredient in the description of 
non-relativistically bound systems like quarkonia. 
In addition it is the basis for the definition of the
lattice coupling as the potential is given by the vacuum expectation
value of the Wilson loop. 

In the perturbative regime
it can be utilized to define a physically motivated strong coupling 
which automatically possesses welcome properties such as gauge invariance
and decoupling of heavy flavors \cite{bmr}. 
The potential can also be employed for a definition of the
coupling from the force between the static sources \cite{gru}.
The latter definition has the advantage that the $\beta$-function is unique
in that no sign-change of the coupling occurs when entering the confinement
regime \cite{gru}.
Moreover, the heavy quark system is ideally suited to
study the overlap region between the non-perturbative and perturbative
treatments of QCD. This latter point is difficult to implement with massless
dynamical fermions and can only be achieved in the coordinate representation.

A possibly very interesting application of the two loop mass corrections to
the heavy quark potential is the effect of a massive charm loop in the
$\overline{\mbox{MS}}$-bottom mass determination \cite{hm}. Using the
physical $\Upsilon$-meson for this purpose, the effect of the mass shift
$\delta m_b$ depends on $\langle \phi_{1s} | V_F(r,m) | \phi_{1s}
\rangle$, where $\phi_{1s}$ denotes the 1s ground state wave function of the
$\Upsilon$-meson and $V_F$ the massive fermionic corrections to the potential.
The effect could be significant if the renormalization scale
depends parametrically on $m_c$, as indicated by a BLM-analysis \cite{hm}.
For a practical evaluation of these mass shifts one needs to have
manageable expressions up to the required order in perturbation theory. The
exact results of Ref. \cite{mel} are not suitable due to the remaining
four Monte Carlo integrations. It might also be easier to perform the mass
shift calculations in position space as only an integration over the
relative distance $r$ would be required. In momentum space, one has an
additional integral as each wave function depends on a different 
three-momentum ${\bf p}$.

In this paper we derive the static QCD-potential in position space with
quark masses through two loops. We use the known results in momentum space
\cite{mel} and derive the coordinate results through a Fourier transformation
of reconstructed approximate 
analytical momentum space expressions. The latter step
is necessary due to the complexity of the results of Ref. \cite{mel}. 

\begin{center}
\begin{figure}
\centering
\epsfig{file=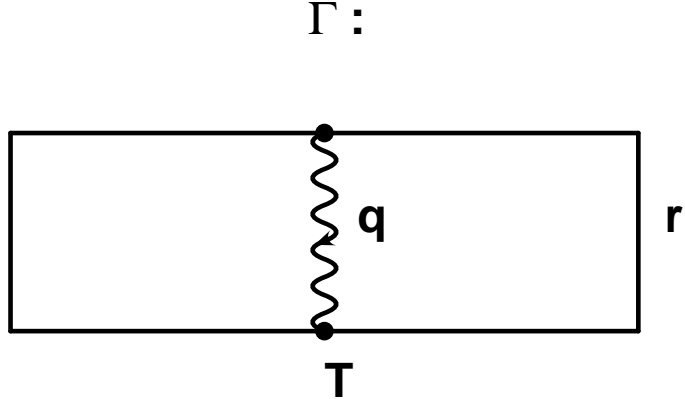,width=3.5in}
\caption{The Wilson-loop $\Gamma$ with large temporal extent ($T \longrightarrow \infty$)
from which the position space potential is defined. Through two loops in
four dimensions, gluons connecting the spatial source lines can be neglected.
} \label{fig:hqpwl}
\end{figure}
\end{center}

\begin{center}
\begin{figure}
\centering
\epsfig{file=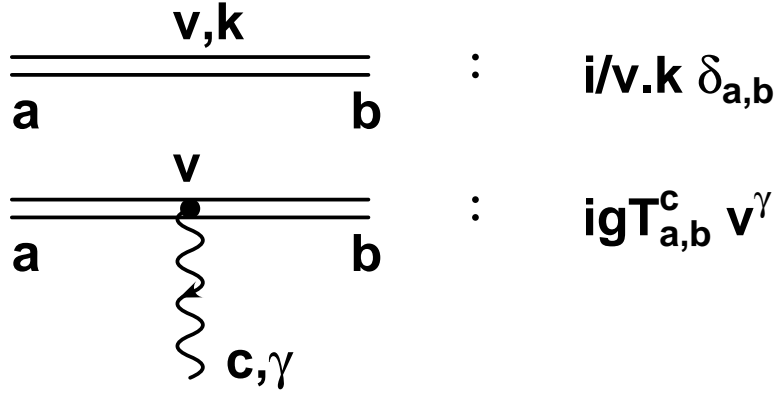,height=2.0in}
\caption{The momentum space Feynman rules used in the calculation of
Ref. \cite{mel}. The $i \varepsilon$-prescription is analogous to the
conventional quark propagator. For anti-sources $v \longrightarrow - v$
must be used.} \label{fig:hqpfr}
\end{figure}
\end{center}

We begin, however, by recalling the definition of the potential through the
manifestly gauge invariant vacuum expectation value of the Wilson loop.
Fig. \ref{fig:hqpwl} displays the Wilson loop $W_\Gamma=\langle 0 | \mbox{Tr}
{\cal P} \exp \left( i g \oint_\Gamma d x_\mu A^\mu_a {\mbox T}^a \right)
|0 \rangle $ of spatial extension
$r$ and large temporal extension $T$ with gluon exchanges indicated. 
The path-ordering is necessary due to the non-commutativity of the SU(3)
generators $\mbox{T}^a$. In
the perturbative analysis through two loops considered here, all spatial
components of the gauge fields $A^\mu_a ({\bf r}, \pm T/2)$ can at most
depend on a power of $\log T$ and are thus negligible here. Furthermore,
$W_\Gamma \stackrel{T \to \infty}{\longrightarrow} \exp \left( - i T E_0 \right)$, where the
ground state energy $E_0$ is identified with the potential $V$. Thus we
arrive at the definition:
\begin{equation}
V(r,m) = - \lim_{T \to \infty} \frac{1}{iT} \log \langle 0| \mbox{Tr}
{\cal P} \exp \left( i g \oint_\Gamma d x_\mu A^\mu_a {\mbox T^a} \right) 
|0 \rangle \label{eq:hqpdef}
\end{equation}
Writing the source term of the heavy charges, separated at the distance
$r \equiv |{\bf r}-{\bf r}^\prime |$, as
\begin{equation}
J^a_\mu (x) = i g v_\mu T^a \left[ \delta ( {\bf x}
- {\bf r} )- \delta ( {\bf x}-{\bf r}^\prime ) \right]
\end{equation}
and neglecting contributions connecting the spatial components,
the perturbative potential is given by
\begin{equation}
V(r,m)=-\lim_{T \to \infty} \frac{1}{iT} \log \langle 0 | \mbox{Tr} {\cal T}
\exp \left( \int d^4x A^a_\mu (x)  J^\mu_a (x) \right) |0 \rangle \label{eq:vj}
\end{equation}
In the above equation
$v_\mu=\delta_{\mu,0}$ due to the purely timelike nature of the sources. For
the same reason the path-ordering is replaced by the time-ordering operator
${\cal T}$.
Expanding Eq. \ref{eq:vj} perturbatively we find the position space Feynman rules
for the source-gluon vertex and the source propagator respectively:
$igT^a v_\mu$ and $-i \theta ( x_0-x_0^\prime) \delta ( {\bf x}- {\bf x}^\prime)
$.
The potential in momentum space is the Fourier transform of $V(r)$. It can be
calculated directly in momentum space from the on-shell heavy quark anti-quark
scattering amplitude (divided by i) at the physical momentum transfer {\bf q},
projected onto the color singlet sector. The momentum space Feynman rule for
the source propagator is identical to the one in HQET \cite{neu}:
$\frac{1}{v \cdot k + i\varepsilon}$. For anti-sources, $v \longrightarrow -v$ is
prescribed (or the time-ordering reversed). Fig. \ref{fig:hqpfr} summarizes
the momentum space Feynman rules. 
In analogy to HQET, double lines denote the heavy
source terms. 

The potential can be used to define the effective charge $\alpha_V$ (
the so-called V-scheme)
through:
\begin{equation}
V(Q,m) \equiv - 4 \pi C_F \frac{\alpha_V (Q,m)}{Q^2} \;\;\;, \;\;\; V(r,m)
\equiv - C_F \frac{\alpha_V(r,m)}{r}
\end{equation}
where $Q^2 \equiv {\bf q}^2=-q^2$ and both expressions above are related
through a Fourier-transform.
The results for QCD-corrections including massless quarks 
have been calculated in Ref. \cite{pet}, and
in Ref. \cite{sch} an independent approach found a disagreement in the pure
glue part of the original results. As both authors agree now on the 
correctness of \cite{sch}, we use in momentum space:
\begin{eqnarray}
\alpha_V(Q) &=& \alpha_{\overline{{\mbox{\tiny MS}}}}(\mu)
\left( 1 + v_1 (Q,\mu) \frac{
\alpha_{\overline{{\mbox{\tiny MS}}}}(\mu)}{\pi} + v_2 (Q,\mu)
\frac{\alpha^2_{\overline{{\mbox{\tiny MS}}}}(\mu)}{\pi^2}\right)
\label{eq:aVm0}
\end{eqnarray}
with
\begin{eqnarray}
v_1(Q,\mu)&=& \frac{1}{4} \left[ \frac{31}{9} C_A - \frac{20}{9} T_Fn_f +
\beta_0 \log \frac{\mu^2}{Q^2} \right] \label{eq:v10} \\
v_2(Q,\mu) &=& \frac{1}{16} \left[ \left( \frac{4343}{162} + 4 \pi -
\frac{\pi^2}{4}
+\frac{22}{3} \zeta_3 \right) C_A^2 - \left( \frac{1798}{81}+ \frac{56}{3} \zeta_3
\right) C_A T_F n_f \right. \nonumber \\ &&
- \left( \frac{55}{3} - 16 \zeta_3 \right) C_F T_F n_f
+ \left( \frac{20}{9} T_F n_f \right)^2 \nonumber \\
&& \left. + \beta_0^2 \log^2 \frac{\mu^2}{Q^2} + \left( \beta_1 + 2 \beta_0 
\left( \frac{31}{9} C_A - \frac{20}{9} T_F n_f \right) \right) \log 
\frac{\mu^2}{Q^2} \right] \label{eq:v20}
\end{eqnarray}
where $\beta_0= \frac{11}{3} C_A - \frac{4}{3} T_F n_f$, $\beta_1
= \frac{34}{3} C_A^2 - \frac{20}{3}C_A T_F n_f - 4 C_F T_F n_f$ and
$C_A=3$, $C_F= \frac{4}{3}$ and $T_F=\frac{1}{2}$ in QCD. The number
of massless flavors is denoted by $n_f$. The $\beta$-function is here
defined as
\begin{equation}
\beta (\alpha_s(\mu^2)) = \frac{1}{\alpha_s(\mu^2)} 
\frac{\partial \alpha_s (\mu^2)}{\partial \log \mu^2}
\equiv - \sum_{n=0}^\infty \beta_n \left( \frac{\alpha_s(\mu^2)}{4 \pi} 
\right)^{n+1} \label{eq:beta}
\end{equation}
For the case of massless quarks, the first two coefficients $\beta_0$ and
$\beta_1$ are renormalization scheme invariant and the $\beta$-function
is gauge invariant to all orders in minimally subtracted schemes \cite{col}.
In coordinate space we have in the massless case \cite{sch2,jps}:
\begin{eqnarray}
\alpha_V(r) &=& \alpha_{\overline{{\mbox{\tiny MS}}}}(\mu)
\left( 1 + v_1 (r,\mu) \frac{
\alpha_{\overline{{\mbox{\tiny MS}}}}(\mu)}{\pi} + v_2 (r,\mu)
\frac{\alpha^2_{\overline{{\mbox{\tiny MS}}}}(\mu)}{\pi^2}\right)
\label{eq:aVr0}
\end{eqnarray}
with
\begin{eqnarray}
v_1(r,\mu)&=& \frac{1}{4} \left[ \frac{31}{9} C_A - \frac{20}{9} T_Fn_f +
2 \beta_0 \log ( \mu r^\prime) \right] \label{eq:v1r0} \\
v_2(r,\mu) &=& \frac{1}{16} \left[ \left( \frac{4343}{162} + 4 \pi -
\frac{\pi^2}{4}
+\frac{22}{3} \zeta_3 \right) C_A^2 - \left( \frac{1798}{81}+ \frac{56}{3} \zeta_3
\right) C_A T_F n_f \right. \nonumber \\ &&
- \left( \frac{55}{3} - 16 \zeta_3 \right) C_F T_F n_f
+ \left( \frac{20}{9} T_F n_f \right)^2 
+ \beta_0^2 \left( 4 \log^2 (\mu r^\prime)  + \frac{ \pi^2}{3} \right)
\nonumber \\
&& \left.
+ 2 \left( \beta_1 + 2 \beta_0 
\left( \frac{31}{9} C_A - \frac{20}{9} T_F n_f \right) \right) \log 
(\mu r^\prime) \right] \label{eq:v2r0}
\end{eqnarray}
where $ r^\prime \equiv r \exp ( \gamma_E )$.

From the renormalon point of view, possible power corrections
in momentum space can at most be of the form\footnote{We still assume that
$Q>\Lambda_{QCD}$ and are concerned only with the form of the leading power
corrections from a renormalon analysis. One cannot learn anything 
about non-perturbative (i.e. confining) effects through this approach.}:
\begin{equation}
V(Q)= - 4 \pi C_F \frac{\alpha_{\overline{{\mbox{\tiny MS}}}}(Q)}{Q^2}
\left( 1 + ... + const. \times \frac{\Lambda^2_{QCD}}{Q^2} + ...
\right) \label{eq:QLQCD}
\end{equation}
which follows from Lorentz invariance (since $v.q=0$) and dimensional 
arguments \cite{ben}.
In this connection it is interesting to note that there is a close
connection between the potential and the pole mass. Both are affected by
the same renormalon ambiguity and writing \cite{ben}:
\begin{equation}
m_{PS}(\mu_f) \equiv m+\frac{1}{2} \int_{Q<\mu_f} \frac{d^3Q}{(2 \pi)^3} V(Q)
\end{equation}
the so-called potential subtracted mass $m_{PS}(\mu_f)$, which depends linearly
on the cutoff $\mu_f$, can be used as a less IR-sensitive mass parameter for
threshold expansions than the pole mass $m$ \cite{ben2}.

In position space, however,
\begin{equation}
V(r)=-C_F \frac{ \alpha_{\overline{{\mbox{\tiny MS}}}}(r)}{r}
\left( 1 + ... + const. \times \Lambda_{QCD} \; r +
const^\prime \times \Lambda^2_{QCD} \; r^2+... \right)\label{eq:rLQCD}
\end{equation}
and it is thus to be expected that the long-distance contributions to the
coordinate space potential are parametrically larger than for the momentum
space potential.
Thus, the position
space potential is likely to be more slowly convergent.
This feature could be interesting when comparing the full lattice results
with the perturbative solution at intermediate distances \cite{bal}.
The form of
the linear term in Eq. \ref{eq:rLQCD} is also motivating the consideration of
the derivative of $V(r)$, i.e. the definition of the strong coupling from
the force between the two heavy quarks \cite{gru}. 

Before considering the effects of massive quarks in the quantum corrections
to the potential, a few general remarks are in order concerning the 
higher order perturbative behavior. The effective theory used for the
calculation amounts to a non-local approach with $Q \sim Mv$, where
$M$ denotes the heavy quark mass, i.e.
where the gluons are always kept off-shell. Through power counting
arguments one can see that through two loops only those gluons need to
be considered. At the three loop level, however, 
on-shell gluon contributions of order $Q \sim Mv^2$
cannot be omitted and a treatment according to the standard definition
would lead to an infrared divergent
potential \cite{bpsv}. These ultra soft terms are expected to be 
canceled, however, by
additional diagrams at that order. In particular, new diagrams connecting
also the spatial components of the Wilson loop in Fig. \ref{fig:hqpwl} will
contribute at three loops. In Ref. \cite{scht} it is indicated how the problem
already shows up at the two loop level in three dimensions and in Ref.
\cite{bpsv} an infrared save definition at higher orders is suggested.
In four dimensions, however, and to the order we are working here
no such problems occur and the definition \ref{eq:vj} is infrared
safe.

The paper is organized as follows. In section \ref{sec:ms} we review the results
of the mass corrections to the heavy quark potential in momentum space at the
two loop order. From the physical Gell-Mann Low equation we reconstruct a simple
analytical approximate expression for the one- and two-loop coefficients.
These results are then used to obtain the position space potential in
section \ref{sec:ps}. In section \ref{sec:af} we then discuss the effect
of massive quarks on the force between two static sources and close with
concluding remarks in section \ref{sec:con}. Explicit results for the
coupling definition through the force in coordinate space are given in
the appendix, where we also discuss briefly the effect of mixed massless and
massive quark loops at the two loop level.

\section{Momentum space results} \label{sec:ms}

\begin{center}
\begin{figure}
\centering
\epsfig{file=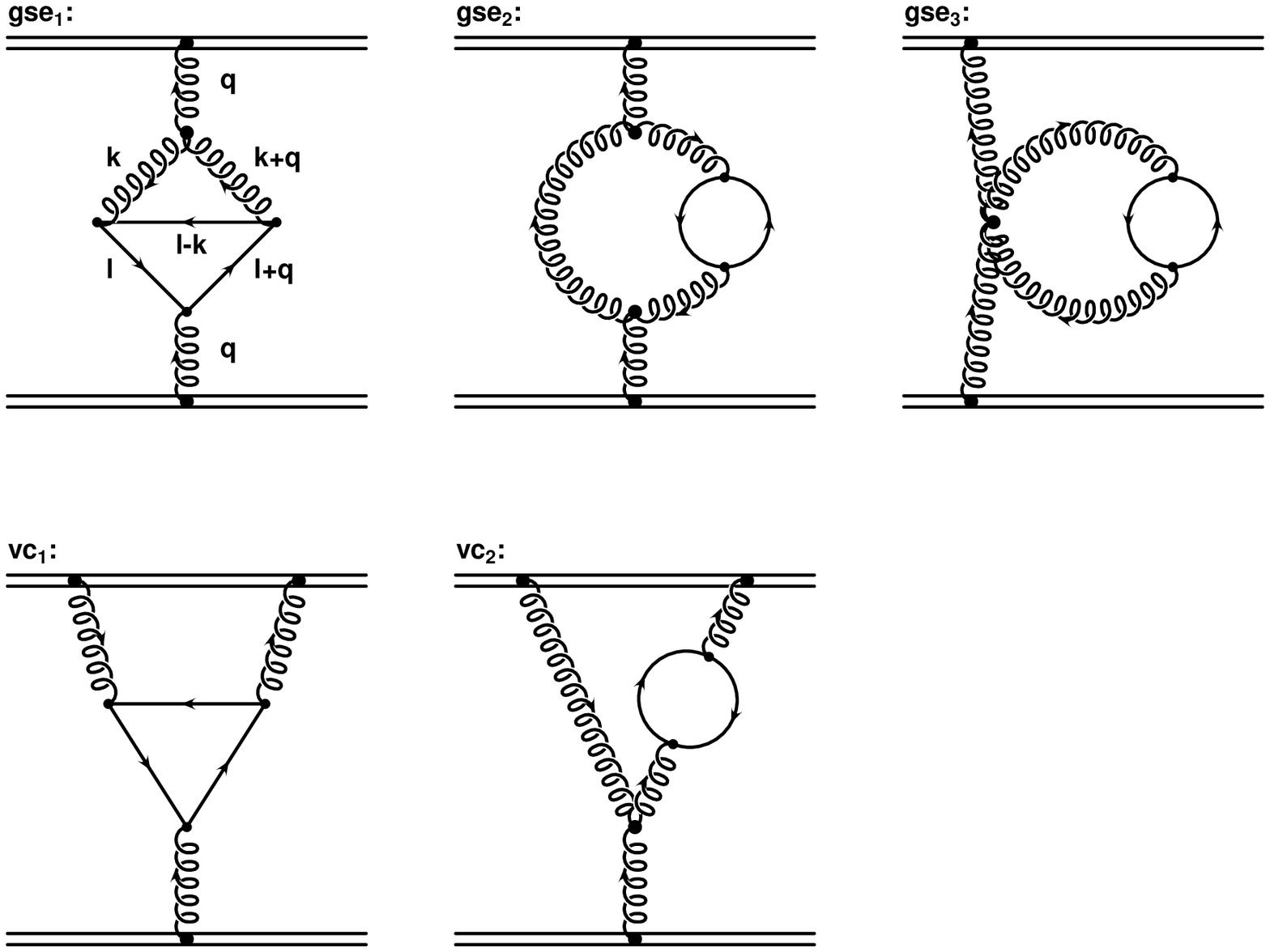,height=2.5in}
\vspace{0.5cm} \\
\epsfig{file=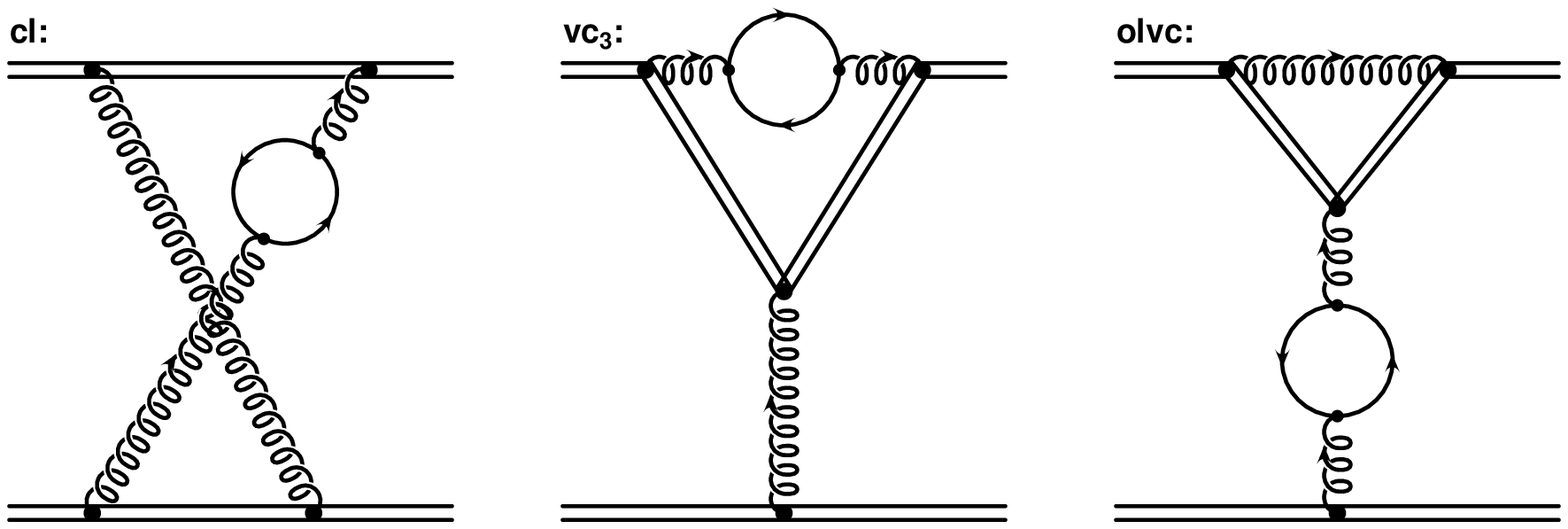,height=1.1in}
\vspace{0.5cm} \\
\epsfig{file=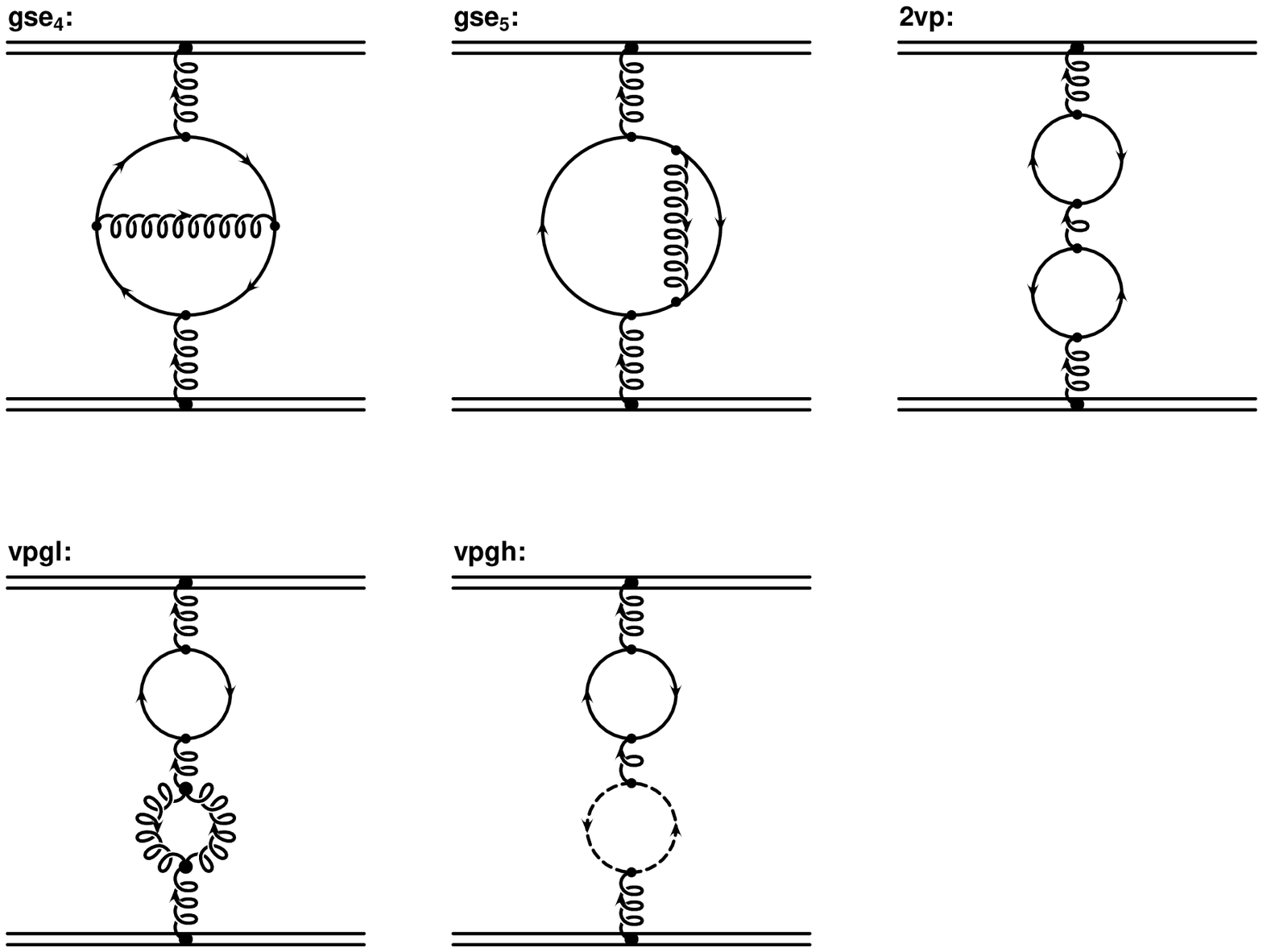,height=2.5in}
\caption{The two-loop massive fermionic corrections to the heavy quark
potential in the Feynman-gauge (from \protect\cite{mel}). 
Double lines denote the heavy
quarks, single
lines the ``light'' quarks with mass $m$. The first two rows contain diagrams
with a typical non-Abelian topology. The middle line includes the infrared
divergent ``Abelian'' Feynman diagrams. They contribute to the potential only
in the non-Abelian theory due to color factors $\propto C_F C_A$. In addition,
although each diagram is infrared divergent, their sum is infrared
finite. The infrared finite Feynman diagrams with an Abelian topology
plus the diagrams consisting of one-loop
insertions with non-Abelian terms are shown in the last two rows.}
\label{fig:tlfd}
\end{figure}
\end{center}

The Monte Carlo results of Ref.~\cite{mel} can be used to obtain the two loop
scale dependence of the static QCD potential in momentum space \cite{bmr}.
The difference to the conventional $\beta$-function in the Callan-Symanzik
\cite{cal,sym} approach is that the physical quark anti-quark system is governed
by the exchanged momentum and independent of the renormalization scale to
each given order. Thus we follow the Gell-Mann Low approach and 
using a running $\overline{\mbox{MS}}$-mass $m(\mu)$ we have:
\begin{eqnarray}
\alpha_V(Q,m(\mu)) &=& \alpha_{\overline{{\mbox{\tiny MS}}}}(\mu)
\left( 1 + v_1 (Q,m(\mu),\mu) \frac{
\alpha_{\overline{{\mbox{\tiny MS}}}}(\mu)}{\pi} + v_2 (Q,m(\mu),\mu)
\frac{\alpha^2_{\overline{{\mbox{\tiny MS}}}}(\mu)}{\pi^2}\right)
\label{eq:aVmu}
\end{eqnarray}
where the massless limit of the
coefficients $v_1$ and $v_2$ is given in Eqs. \ref{eq:v10} and \ref{eq:v20}.
The physical charge $\alpha_V(Q,m(\mu))$ cannot depend explicitly on the
renormalization scale $\mu$ and
the explicit $\mu$-dependence on the right-hand side of Eq.~(\ref{eq:aVmu})
cancels to the order we are working.
Fig.~\ref{fig:tlfd} gives the Feynman diagrams
for the fermionic contributions to the two-loop coefficient $v_2(Q,m(\mu),\mu)$.
The mass-counterterm chosen for the Feynman diagram labeled $gse_5$ determines
the mass-parameter which has to be used in 
the one-loop coefficient $v_1(Q,m(\mu),\mu)$. 
In Ref. \cite{bmr} we considered the flavor-threshold dependence of heavy
quarks and related the running mass to the pole mass which is
renormalization-scale independent and gives explicit decoupling. This also
provides a physical picture as well as a straightforward Abelian
limit.
The relation between the $\overline{\mbox{MS}}$ mass $m(\mu)$ and the pole mass $m$
is given by \cite{tar},
\begin{equation}
m(\mu) = m \left[ 1 - C_F \frac{\alpha_{\overline{{\mbox{\tiny MS}}}}(\mu)}{\pi}
\left( 1 + \frac{3}{2} \log \frac{
\mu}{m} \right) \right],
\label{eq:mrm}
\end{equation}
Inserting Eq.~(\ref{eq:mrm}) into  Eq.~(\ref{eq:aVmu}) gives
\begin{eqnarray}
\alpha_V(Q,m) &=& \alpha_{\overline{{\mbox{\tiny MS}}}}(\mu)
\left[ 1 +
v_1 (Q,m,\mu) \frac{\alpha_{\overline{{\mbox{\tiny MS}}}}(\mu)}{\pi} + \right. \nonumber \\
&& \left. \left[v_2 (Q,m,\mu) + \Delta_m(Q,m,\mu)\right]
\frac{\alpha_{\overline{\mbox{\tiny MS}}}^2(\mu)}{\pi^2}
\right] \label{eq:aVmupm}
\end{eqnarray}
where $\Delta_m(Q,m,\mu)$ denotes the contribution arising from $v_1$
when changing from the running mass to the pole mass:
$v_1(Q,m(\mu),\mu) = v_1(Q,m,\mu) + \Delta_m(Q,m,\mu)
\frac{\alpha_{\overline{{\mbox{\tiny MS}}}}(\mu)}{\pi}.$

The Gell-Mann Low function \cite{gl} for the $V$-scheme is defined as the
total logarithmic
derivative of the effective charge with respect to the physical
momentum transfer scale $Q$:
\begin{equation}\label{eq:psiv}
\Psi_V \left( \frac{Q}{m} \right) \equiv  \frac{d \alpha_V (Q,m)}{
d \log Q} \equiv \sum^{\infty}_{i=0} -\psi_{V}^{(i)} \frac{\alpha_V^{i+2} (Q,m)}
{\pi^{i+1}} \; ,
\end{equation}
where in the massless case the coefficients $\psi_{V}^{(0)}$ and
$\psi_{V}^{(1)}$ are given by,
\begin{eqnarray}
\psi_{V}^{(0)}\left(m=0 \right) &=&  \frac{11}{6}C_A-\frac{2}{3}T_F N_F =
\frac{11}{2}-\frac{1}{3}N_F \; ,\\
\psi_{V}^{(1)} \left( m=0 \right)&=&
\frac{17}{12}C_A^2-\frac{5}{6}C_AT_FN_F-\frac{1}{2}C_FT_FN_F
= \frac{51}{4} -\frac{19}{12}N_F \; .
\end{eqnarray}
For the massive case all the mass effects are absorbed into a
mass-dependent function $N_F$. In other words we write
\begin{eqnarray}\label{eq:nfv0}
\psi_{V}^{(0)}\left( \frac{Q}{m}  \right) &=&
\frac{11}{2}-\frac{1}{3}N_{F,V}^{(0)}\left( \frac{Q}{m} \right) \\
\label{eq:nfv1}
\psi_{V}^{(1)} \left(  \frac{Q}{m}  \right)&=&
 \frac{51}{4} -\frac{19}{12}N_{F,V}^{(1)}\left( \frac{Q}{m} \right) \; ,
 \end{eqnarray}
 where the subscript $V$ indicates the scheme dependence of
 $N_{F,V}^{(0)}$ and $N_{F,V}^{(1)}$.

 Taking the derivative of  Eq.~(\ref{eq:aVmupm}) with respect to $\log Q$ and
 re-expanding the result in  $\alpha_V(Q,m)$ gives the following equations
 for the first two coefficients of $\Psi_V$:
 \begin{eqnarray}
 \psi_{V}^{(0)}\left( \frac{Q}{m} \right) &=& -\frac{d v_1 (Q,m,\mu)}
 {d \log Q} \label{eq:psi0} \\
 \psi_{V}^{(1)} \left( \frac{Q}{m} \right)&=&
 -\frac{d [v_2 (Q,m,\mu)+ \Delta_m(Q,m,\mu)]}{d \log Q} + 2
 v_1 (Q,m,\mu) \frac{d v_1 (Q,m,\mu)}{d \log Q} \; .
 \label{eq:psi1}
 \end{eqnarray}
 The argument $Q/m$  indicates that there is no
 renormalization-scale  dependence in Eqs.~(\ref{eq:psi0}) and
 (\ref{eq:psi1}). Rather, $\psi_{V}^{(0)}$ and $\psi_{V}^{(1)}$ are functions
 of the ratio of the physical momentum transfer
 $Q = \sqrt{-q^2}$ and the pole mass $m$ only.
 The expression for $\psi_{V}^{(0)}$ agrees
 with our result in Ref.~\cite{bgmr}.
 In Eq.~(\ref{eq:psi1}) the derivative of the $\Delta_m(Q,m,\mu)$-term comes
 from using the pole-mass instead of the MS mass whereas
 the remaining mass dependence in  Eq.~(\ref{eq:psi1})  is arbitrary
 in the sense that a different mass scheme is formally of higher order.
 In addition we note that
 the contribution $2v_1 dv_1/d \log Q$ cancels the reducible
 contribution (labeled  {\bf 2vp} in Fig.~\ref{fig:tlfd}) to
 $v_2$;  it is thus sufficient to consider one quark flavor
 at a time for the two loop Gell-Mann Low function. 
 In the appendix we describe how to treat the effect of massless
 quark loops (u,d and s) in mixed Feynman diagrams on the amplitude level
 as well as the case for two different mass flavor loops.

Because of the complexity of the integrals encountered in the
evaluation~\cite{mel} of the massive two-loop corrections to the
heavy quark potential, the  results were obtained
numerically using the adaptive Monte Carlo integration program
VEGAS~\cite{veg}. Thus the derivative of the two-loop term
$v_2$ was calculated numerically, whereas the other terms in
Eqs.~(\ref{eq:psi0}) and (\ref{eq:psi1}) were obtained analytically.
The results are given in terms
of the contribution to the effective number of flavors
$N^{(0)}_{F,V}\left( \frac{Q}{m} \right)$ and
$N^{(1)}_{F,V}\left( \frac{Q}{m} \right)$ in the $V$-scheme
from a given quark with mass
$m$ defined according to Eqs.~(\ref{eq:nfv0}) and (\ref{eq:nfv1})
respectively. The Appelquist-Carazzone~\cite{ac}  theorem
requires the decoupling of heavy masses at small momentum
transfer for physical observables. Thus 
$N^{(1)}_{F,V} \left(
\frac{Q}{m} \right)$ goes to zero for  $Q/m \to 0$.  The massless result
$N_{F,V}^{(1)} \to 1$ is also be recovered for large scales.

The calculation presented in Ref.~\cite{mel} required the evaluation
of four-dimensional scalar integrals. The results in Ref. \cite{bmr}
are based on
 50 iterations of the integration grid each comprising
 $10^7$ evaluations of the function which where needed to achieve
 adequate convergence.  Even so,
 the Monte Carlo results still are not completely stable for  small
 values of $Q/m$, especially in the light of the numerical differentiation
 required in Eq.~(\ref{eq:psi1}).  Nevertheless,
 accurate results can be obtained by fitting the numerical
 calculation to a suitable analytic function.

 The one-loop contribution to the effective number of flavors $N_F$
 follows from the standard  formula for QED vacuum polarization.
 In Ref.~\cite{bgmr}
 we used the simple representation in terms of a rational
 polynomial~\cite{rg}:
 \begin{equation}
 N^{(0)}_{F,V} \left( \frac{Q}{m} \right) \approx \frac{1}{1+5.2 \frac{m^2}{Q^2}} \equiv \frac{1}{1+a_0 \frac{m^2}{Q^2}}
 \label{eq:nf0}
 \end{equation}
 which displays decoupling  for small
 scales and the correct massless limit at large
 scales.  Similarly,  the numerical results for the two-loop contribution
 can be fit to the form
 \begin{equation}
 N^{(1)}_{F,V} \left( \frac{Q}{m} \right) \approx
 \frac{a_1 \displaystyle\frac{Q^2}{m^2}+a_2 \displaystyle\frac{Q^4}{m^4}}
 {1+a_3 \displaystyle\frac{Q^2}{m^2} + a_2\displaystyle\frac{Q^4}{m^4}}
 \label{eq:psi1fit}
 \end{equation}
 The parameter values $a_i$ and the errors obtained from the fit to the
 numerical calculation in the $V$-scheme for QCD and QED are given in Ref.
 \cite{bmr}. Similar decoupling forms have been used for
 interpolating the flavor dependence of the effective coupling in the
 momentum subtraction schemes (MOM)~\cite{yh,jt}.

\begin{figure}[htbp]
 \center
 \epsfig{file=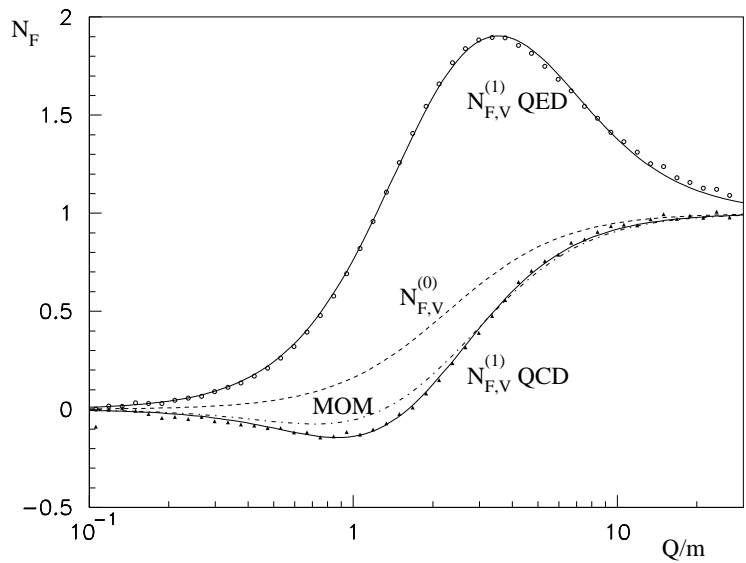,height=3.0in}
 \caption{The numerical results for the gauge-invariant  $N_{F,V}^{(1)}$ in QED
 (open circles) and QCD (triangles) with the best $\chi^2$ fits of
 Eqs.~(\ref{eq:psi1qed}) and (\ref{eq:psi1qcd})
 superimposed respectively (from Ref. \cite{bmr}).   The dashed
 line shows the one-loop $N_{F,V}^{(0)}$ function  of Eq.~(\ref{eq:nf0}).
 For comparison we
 also show the gauge dependent two-loop result obtained in MOM schemes
 (dash-dot)  \protect\cite{yh,jt}. At large $\frac{Q}{m}$ the theory becomes
 effectively massless, and both schemes agree as expected. The figure also
 illustrates the decoupling of heavy quarks at small $\frac{Q}{m}$.}
 \label{fig:nfV}
 \end{figure}

 In the case of QCD we obtain the following approximate form for the
 effective number of flavors for a given quark with pole-mass $m$:
 \begin{equation}
 N^{(1)}_{F,V} \left( \frac{Q}{m} \right) \approx \frac{\left( -0.571 + 0.221
 \displaystyle\frac{Q^2}{m^2} \right)
 \displaystyle\frac{Q^2}{m^2}}{1+1.326 \displaystyle\frac{Q^2}{m^2} + 0.221
 \displaystyle\frac{Q^4}{m^4}}
 \label{eq:psi1qcd}
 \end{equation}
 and for QED
 \begin{equation}
 N^{(1)}_{F,V} \left( \frac{Q}{m} \right) \approx \frac{\left( 1.069 + 0.0133
 \displaystyle\frac{Q^2}{m^2} \right)
 \displaystyle\frac{Q^2}{m^2}}{1+0.402 \displaystyle\frac{Q^2}{m^2} + 0.0133
 \displaystyle\frac{Q^4}{m^4}} \; .
 \label{eq:psi1qed}
 \end{equation}
 The results of our numerical calculation of $N_{F,V}^{(1)}$ in the
 $V$-scheme for QCD and QED are shown in Fig.~\ref{fig:nfV}.
 The decoupling of heavy quarks becomes manifest at small $Q/m$, and
 the massless limit is attained for large $Q/m$. The QCD form actually
 becomes negative at moderate values of $Q/m$, a novel feature of
 the anti-screening non-Abelian contributions. This property is
 also present in the (gauge dependent) MOM results.
 In contrast, in Abelian QED the two-loop contribution to
 the effective number of flavors becomes larger than 1 at intermediate
 values of $Q/m$.  We also display the one-loop
 contribution $N^{(0)}_{F,V} \left( \frac{Q}{m} \right)$ which
 monotonically  interpolates between the decoupling and massless limits.
 The solid curves displayed in
 Fig.~\ref{fig:nfV} show that the parameterizations of
 Eq.~(\ref{eq:psi1qcd})  which we used for fitting the numerical results are
 quite accurate. 

In Ref. \cite{bmr} it was shown that the Abelian limit displayed in Fig.
\ref{fig:nfV} agrees with the well known literature results \cite{ks,br,kni} and
that the full QCD result is independent of the renormalization scale.
The very good agreement of the exact two loop calculation with the
relatively simple fitting function for $N_{F,V}^{(1)}$ of Eq.~\ref{eq:psi1qcd}
makes it possible to reconstruct an analytical approximate function for the
full mass dependent two loop coefficient in the next section. 

\subsection{Reconstructing the momentum space potential} \label{sec:rc}

Starting from the general expression for the Gell-Mann Low function in Eq.
\ref{eq:psiv} we can obtain $\alpha_V(Q,m)$ through integration over $\log Q$.
Our goal is to reconstruct an analytical function for $\alpha_V(Q,m)$ in
terms of $\alpha_{\overline{\mbox{\tiny MS}}}(\mu)$ based on the fitting parameters
$a_0$ and $a_1,a_2,a_3$ from the approximate one- and two-loop solutions
respectively. It should be clear that by analytical we mean an expression
of known functions depending on $\{Q^2,m^2,\mu^2,a_0,a_1,a_2,a_3\}$ in the
spacelike (Euclidean) region. It is understood that for a continuation into
the timelike regime only the full function calculated in Ref. \cite{mel}
can be used, not the approximate
form we derive below. With this in mind we can write
\begin{eqnarray}
\alpha_V(Q,m)-\alpha_{\overline{\mbox{\tiny MS}}}(\mu) &=& -\frac{\alpha^2_{\overline{\mbox{\tiny MS}}}
(\mu)}{\pi} \left( \int \psi^{(0)}_V \;\; d \, \log Q + C^{(0)} \right) \nonumber
\\ && -\frac{\alpha^3_{\overline{\mbox{\tiny MS}}}(\mu)}{\pi^2} \left( \int 
\left[ \psi^{(1)}_V 
-2 v_1 \psi^{(0)} \right] d \, \log Q + C^{(1)} \right)
\end{eqnarray}
The integration constants $C^{(i)}$ can be functions of $m$ and $\mu$ and are
fixed by requiring that the correct massless limit is obtained.
We find for the corrections with one massive quark with pole-mass $m$:
\begin{eqnarray}
v_1(Q,m,\mu) &=& \frac{31}{36} C_A - \frac{11}{12} C_A \log \frac{
Q^2}{\mu^2} + \frac{T_F}{9} \left( -5 + 3 \log \frac{Q^2+a_0m^2}{\mu^2} \right)
\nonumber \\
v_2(Q,m,\mu) &=& \frac{C^2_A}{16} \left( \frac{4343}{162} + 4 \pi^2-
\frac{\pi^4}{4} + \frac{22}{3} \zeta_3 +\frac{121}{9} \log^2 \frac{Q^2}{\mu^2}
- \frac{988}{27} \log \frac{Q^2}{\mu^2} \right) \nonumber \\
&& + T_F \left[ -\frac{C_A}{16} \left( \frac{1798}{81}+\frac{56}{3} \zeta_3 
\right) - \frac{C_F}{16} \left( \frac{55}{3} - 16 \zeta_3 \right) + \frac{25}{
81} T_F \right. \nonumber \\
&&+\frac{19}{6} \left( \frac{1}{4} \log \frac{m^4+a_3m^2Q^2+a_2Q^4}{a_2\mu^4}
+\frac{\frac{a_3}{4}-\frac{a_1}{2}}{\sqrt{a_3^2-4a_2}} 
\times \right. \nonumber \\ && \left. \log
\frac{2a_2Q^2+(\sqrt{a_3^2-4a_2}+a_3)m^2}{2a_2Q^2-(\sqrt{a_3^2-4a_2}-a_3)m^2}
\right) +\frac{55}{54} C_A \log \frac{Q^2}{\mu^2} + \nonumber \\ &&
\frac{31}{54} C_A
\log \frac{Q^2+a_0m^2}{\mu^2} -\frac{10}{27}T_F \log \frac{Q^2+a_0m^2}{\mu^2}
-\frac{11}{36} C_A \log^2 \frac{Q^2}{\mu^2} 
\nonumber \\
&& \left. -\frac{11}{18} C_A \log 
\frac{Q^2}{\mu^2} \log \frac{Q^2+a_0m^2}{Q\mu}
+ \frac{1}{9} T_F \log^2 \frac{Q^2+a_0m^2}{\mu^2} \right] \label{eq:aVm}
\end{eqnarray}
In terms of the running mass in the $\overline{\mbox{MS}}$-scheme the solution
reads:
\begin{equation}
v_2(Q,m(\mu),\mu)=v_2(Q,m,\mu)
+ \frac{1}{6}
C_F T_F \frac{4+3 \log \frac{\mu^2}{m^2}}{1+ \frac{Q^2}{a_0m^2}} \label{eq:aVmsb}
\end{equation}
The results are written in such a form that the limit $m \longrightarrow 0$
is obvious and can be seen to agree with Eq. \ref{eq:aVm0}. Eq. \ref{eq:aVmsb}
can be directly compared to the exact calculation in Ref. \cite{mel}. The latter
was obtained in the MS-scheme, so we must use
\begin{equation}
\mu_{\tiny{\mbox{MS}}}=\sqrt{\frac{e^{\gamma_E}}{4\pi}} \mu_{\overline{
\tiny{\mbox{MS}}}}
\end{equation}
 \begin{figure}
 \center
 \epsfig{file=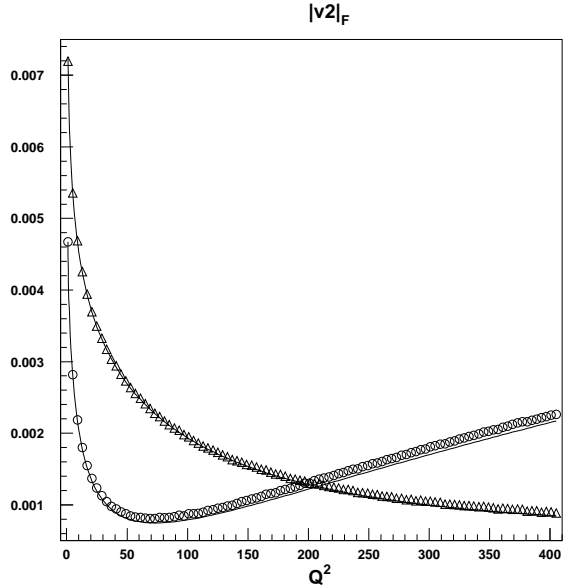,height=3.5in}
 \caption{The comparison between the exact results from Ref. \cite{mel} (open
 symbols) and the reconstructed solution in Eq. \ref{eq:aVmsb} (solid lines) 
 for the
 bottom (triangles) and charm (circles) quarks. 
 The absolute value of the fermionic contributions
 (proportional to $T_F$) times $\frac{1}{16 \pi^4}$ is shown.
 The scale $\mu$ was
 chosen to coincide with the quark masses in the MS-scheme of Ref. \cite{mel}.
 It can clearly be seen that Eq. \ref{eq:aVmsb} is in good agreement with
 the full analytical result over all perturbative values of the momentum
 transfer $Q$ within the statistical Monte Carlo and fitting
 errors of a few percent in each case.}
 \label{fig:TF}
 \end{figure}
Fig. \ref{fig:TF} shows the good agreement of approximate solution in
Eq. \ref{eq:aVmsb} with the full result for different input parameters.
The figure also displays the fact that the fitting parameters $a_0,..,a_3$
of Ref. \cite{bmr} are optimized for the flavor threshold
region $Q\sim {\cal O} (m)$.
 \begin{figure}
 \center
 \epsfig{file=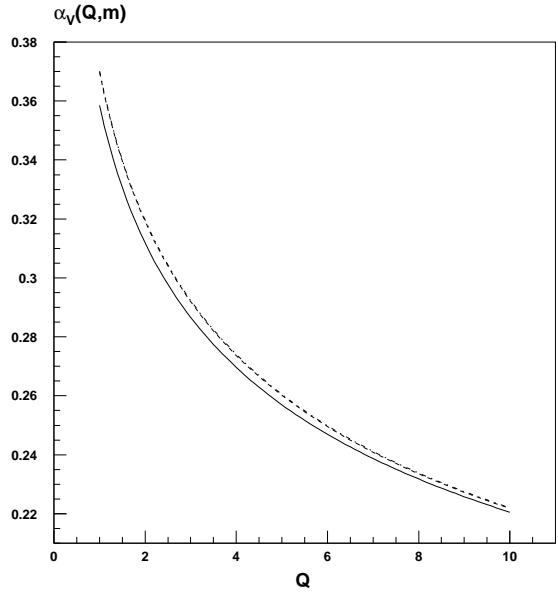,width=3.5in}
 \caption{The momentum dependence of $\alpha_V(Q,m)$. We use 
 $\mu=M_Z$, $\alpha_{\overline{\mbox{\tiny MS}}}(M_Z)=0.12$ and $n_f=5$
 throughout. The solid line is the massless result, while the two dashed
 lines on top of each other
 are massive results for the b-quark with the pole and $\overline
 {\mbox{MS}}$ mass.}
 \label{fig:aVQ}
 \end{figure}
In Fig. \ref{fig:aVQ} the two loop $Q$-dependence of the physical $\alpha_V(Q,m)$
charge is compared with the massless calculation. In the massive
result the bottom quark mass $m_b=4.5$GeV is used and it can be seen that
the two massive results (pole and running mass) differ by several percent
from the massless ($n_f=1$) calculation in the region $Q\sim m_b$.
It should also be stressed that we display the full $Q$-behavior of
$\alpha_V(Q,m)$ including constants, 
not just the running according to the Gell-Mann Low function.

\section{Coordinate space results} \label{sec:ps}

In this section we will present results for the coordinate space potential
based on the Fourier transform of Eq. \ref{eq:aVmsb}. It should be
emphasized that the Fourier transform of $\alpha_V(Q,m)$ in the strictly
perturbative sense does not exist (Landau-pole) and that only the expanded
coefficients can be used. This will be shown below.
In general, we have the 
following relations:
\begin{eqnarray}
V(r,m)&=& - C_F \frac{\alpha_V(r,m)}{r}=\int \frac{d^3Q}{(2\pi)^3}
V(Q,m) \exp ( i {\bf Qr}) \nonumber \\
&=&-4 \pi C_F \int \frac{d^3Q}{(2\pi)^3}
\frac{\alpha_V(Q,m)}{Q^2} \exp ( i {\bf Qr}) \nonumber \\
&=& - \frac{2}{\pi} \; \frac{C_F}{r} \int^\infty_0 \frac{dQ}{Q} \sin (Qr)
\alpha_V(Q,m) \label{eq:Vrm}
\end{eqnarray}
At fixed orders, there is no Landau-pole in Eq. \ref{eq:Vrm} which can best
be seen by inserting the expansion in Eq. \ref{eq:aVmu} into Eq. \ref{eq:Vrm}:
\begin{eqnarray}
V(r,m)&=&- \frac{2}{\pi} \; \frac{C_F \alpha_{\overline{\mbox{\tiny MS}}}(\mu)}{r}
\int^\infty_0 \frac{dQ}{Q} \sin (Qr) \left(1 + v_1(Q,m,\mu) 
\frac{\alpha_{\overline{\mbox{\tiny MS}}}(\mu)}{\pi} \right. \nonumber \\ && + 
\left. v_2(Q,m,\mu) 
\left( \frac{\alpha_{\overline{\mbox{\tiny MS}}}(\mu)}{\pi} \right)^2 \right)
\end{eqnarray}
It can easily be recognized that the first term in the series is just the
Coulomb potential $V_C=-C_F \frac{\alpha_{\overline{\mbox{\tiny MS}}}(\mu)}{r}$.
The loop corrections are then given by writing
\begin{equation}
\alpha_V(r,m)=\alpha_{\overline{\mbox{\tiny MS}}} (\mu) \left( 1 +
v_1(r,m,\mu) \frac{\alpha_{\overline{\mbox{\tiny MS}}}(\mu)}{\pi}
+v_2(r,m,\mu) \left( \frac{
\alpha_{\overline{\mbox{\tiny MS}}}(\mu)}{\pi}
\right)^2 \right) \label{eq:aVrm}
\end{equation}
with 
\begin{eqnarray}
v_1(r,m,\mu)&=&\frac{C_A}{4} \left( \frac{31}{9} + \frac{22}{3} \left( \log (\mu
\; r)
+ \gamma_E \right) \right) - \frac{5}{9}T_F \nonumber \\ 
&& + \frac{T_F}{3} \left( \log \frac{a_0 m^2}{\mu^2} + 2 \mbox{Ei} (1,\sqrt{a_0}\; m \;r) 
\right) \\      
v_2(r,m,\mu) &=& \frac{C^2_A}{16} \left( \frac{4343}{162} + 4 \pi^2-
\frac{\pi^4}{4} + \frac{22}{3} \zeta_3 +\frac{121}{9} \left[ 4 \left( \log 
(\mu \; r) + \gamma_E \right)^2+\frac{\pi^2}{3} \right] \right. \nonumber \\ &&
\left. + \frac{1976}{27} \left( \log (\mu \; r) + \gamma_E \right) \right) 
\nonumber \\
&& + T_F \left[ -\frac{C_A}{16} \left( \frac{1798}{81}+\frac{56}{3} \zeta_3 
\right) - \frac{C_F}{16} \left( \frac{55}{3} - 16 \zeta_3 \right) + \frac{25}{81}
T_F \right. \nonumber \\
&& + \frac{19}{6} \left\{ \frac{1}{4} \log \frac{(a_3+\sqrt{a_3^2-4a_2})m^2}{2
a_2\mu^2}+ \frac{1}{2} \mbox{Ei} \left( 1, \sqrt{\frac{a_3+\sqrt{a_3^2-4a_2}}{2a_2}} \;
m \; r \right)
\right. \nonumber \\ && 
+ \frac{1}{4} \log \frac{(a_3-\sqrt{a_3^2-4a_2})m^2}{2a_2\mu^2}+ \frac{1}{2}
\mbox{Ei} \left( 1,
\sqrt{\frac{a_3-\sqrt{a_3^2-4a_2}}{2a_2}} \; m \; r \right) \nonumber \\
&& \left. + \frac{ \frac{a_3}{4}-\frac{a_1}{2}}{\sqrt{a^2_3-4a_2}} \left( \log 
\frac{ a_3+\sqrt{a_3^2-4a_2}}{a_3-\sqrt{a_3^2-4a_2}} +2 \;
\mbox{Ei} \left( 1, \sqrt{a_3+\sqrt{a_3^2-4a_2}} \; m \; r \right) \right. \right. \nonumber \\
&& \left. \left. - 2 \; \mbox{Ei} \left( 1, \sqrt{
a_3-\sqrt{a_3^2-4a_2}} \; m \; r \right) \right) \right\} - \frac{55}{27} C_A \left(
\log (\mu \; r) + \gamma_E \right) \nonumber \\ &&
+ \left( \frac{31}{54} C_A - \frac{10}{27} T_F \right) \left( \log \frac{a_0m^2}{
\mu^2} + 2 \mbox{Ei} \left( 1, \sqrt{a_0} \; m \;r \right) \right)- \frac{11}{36} C_A
\left( \frac{\pi^2}{3} \right. \nonumber \\ &&
\left. \left. + 4 \left( \log (\mu \; r) + \gamma_E \right)^2  
\right) - \frac{ 11}{18} C_A \frac{2}{\pi} {\cal I}_1 (r,m,\mu) + \frac{T_F}{9} 
\frac{2}{\pi} {\cal I}_2 (r,m,\mu) \right]
\end{eqnarray}
where the integral representations are defined as follows:
\begin{eqnarray}
\mbox{Ei} (1,x) &\equiv& \int^\infty_x \exp (-t) \frac{dt}{t} \label{eq:ei} \\
{\cal I}_1 (r,m,\mu) &\equiv& \int^\infty_0 \log \frac{ Q^2+a_0m^2}{Q\mu} \log \frac{Q^2}{
\mu^2} \sin ( Q \; r) \frac{dQ}{Q} \label{eq:ir1} \\
{\cal I}_2 (r,m,\mu) &\equiv& \int^\infty_0 \log^2 \frac{ Q^2+a_0m^2}{\mu^2}
\sin ( Q \; r) \frac{dQ}{Q} \label{eq:ir2}
\end{eqnarray}
Eqs. \ref{eq:ei}, \ref{eq:ir1} and \ref{eq:ir2} are readily calculable numerically, for instance with the mathematical package
MATHEMATICA\footnote{For Eqs. \ref{eq:ir1} and 
\ref{eq:ir2} the integration method should be adapted to the oscillatory
behavior of the integrands.}.
Again, if one uses the $\overline{\mbox{MS}}$-running mass $m(\mu)$ instead of
the above used pole mass $m$ the result must be modified in the following
way:
\begin{equation}
v_2(r,m(\mu),\mu)=v_2(r,m,\mu)+\frac{1}{6} C_F T_F \left( 4 + 3 \log \frac{\mu^2}{
m^2} \right) \left( 1 - \exp (- \sqrt{a_0} \;m \;r ) \right)
\end{equation}
 \begin{figure}
 \center
 \epsfig{file=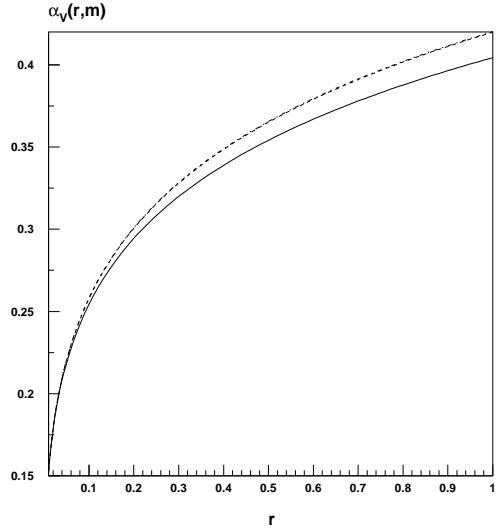,height=3.1in}
 \caption{The distance dependence of $\alpha_V(r,m)$ for the same parameters
 as in Fig. \ref{fig:aVQ}. $r$ is in GeV$^{-1}$ and $0.2 \mbox{fm} \approx 1$GeV$^{-1}$.
 }
 \label{fig:aVr}
 \end{figure}
 \begin{figure}
 \center
 \epsfig{file=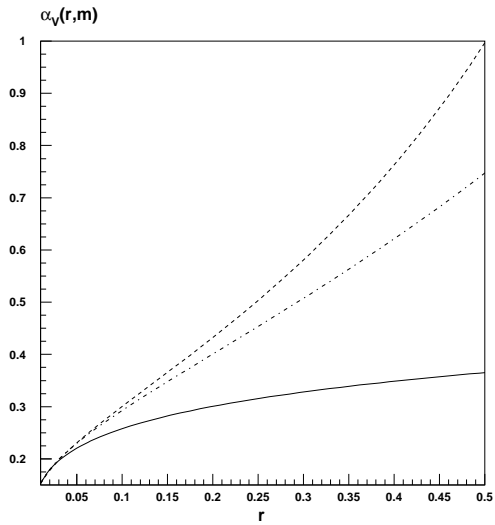,height=3.1in}
 \caption{The distance dependence of $\alpha_V(r,m)$ for different
 choices of the renormalization scale $\mu$. The solid line corresponds
 to $\mu=M_Z$ as in Fig. \ref{fig:aVr}. The dash-dotted line to
 $\mu=\frac{1}{r}$ and the dashed line to the ``natural choice''
 $\mu=\frac{1}{r e^{\gamma_E}}$. The source separation $r$ is in GeV$^{-1}$.}
 \label{fig:aVmu}
 \end{figure}
The effect of the two loop mass effects in position space are depicted in
Fig. \ref{fig:aVr} for the bottom quark with $m_b=4.5$GeV. The pole-
and $\overline{\mbox{MS}}$-mass corrections are almost identical on
the scale (dashed lines) and the effect grows with larger distances.
Since in the renormalization group (RG) logarithms always enter as $\log (\mu r)
+\gamma_E$,
the ``natural" choice is given by $\mu=\frac{1}{r e^{\gamma_E}}$. In Ref.
\cite{pett} it was shown that this scale almost identically reproduces the
BLM-results \cite{blm} in the massless case, thus consistently reabsorbing the large
RG-group effects. As in this case we also have to evaluate the two-loop running
of the $\overline{\mbox{MS}}$-coupling using:
\begin{equation}
\alpha_{\overline{\mbox{\tiny MS}}} (\mu) = \frac{4 \pi}{\beta_0 \log
\frac{\mu^2}{\Lambda^2_{QCD}}} \left(1 - \frac{\beta_1}{\beta_0^2} 
\frac{ \log \left( \log \frac{\mu^2}{\Lambda^2_{QCD}} \right)}{\log
\frac{\mu^2}{\Lambda^2_{QCD}}} \right) \label{eq:Lqcd}
\end{equation}
where we normalize the QCD-scale parameter $\Lambda_{QCD}$ such that
$\alpha_{\overline{\mbox{\tiny MS}}} (M_Z) = 0.12$ which corresponds to
$\Lambda_{QCD}=0.25$GeV and we keep $n_f=5$ fixed. Fig. \ref{fig:aVmu}
displays the effect of the various scale choices on $\alpha_V(r,m)$.
Although the physical coupling $\alpha_V(r,m)$ has no renormalization
scale dependence to the order we are working, the difference seen in
Fig. \ref{fig:aVmu} is due to uncanceled higher order terms and, in 
general, increases at larger distances.
Substantial deviations can be seen for $\alpha_V(r,m)>0.2$.
 \begin{figure}
 \center
 \epsfig{file=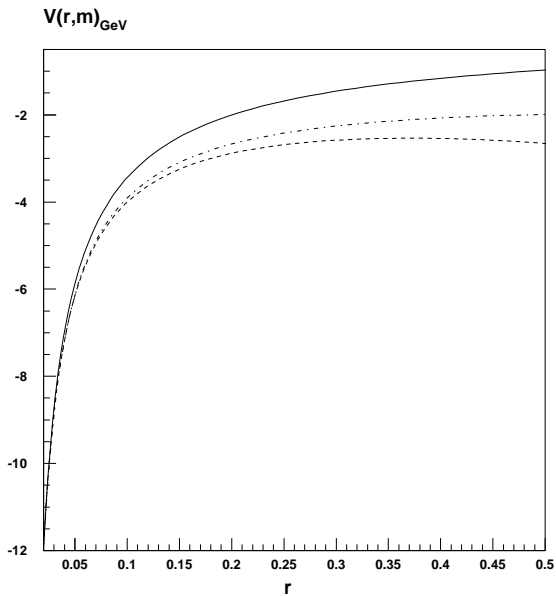,height=3.5in}
 \caption{The distance dependence of $V(r,m)$ with the same
 choices of the renormalization scale $\mu$ as in Fig. \ref{fig:aVmu}.
 The potential is given in units of GeV, $r$ in GeV$^{-1}$.}
 \label{fig:Vmu}
 \end{figure}
The potential is shown in Fig. \ref{fig:Vmu} and one has to use the standard 
conversion factor $1 \mbox{fm}^{-1} \approx 0.2$GeV in order to convert
the distance to fm. Again, the scale dependence is due to higher order
contributions.
 \begin{figure}
 \center
 \epsfig{file=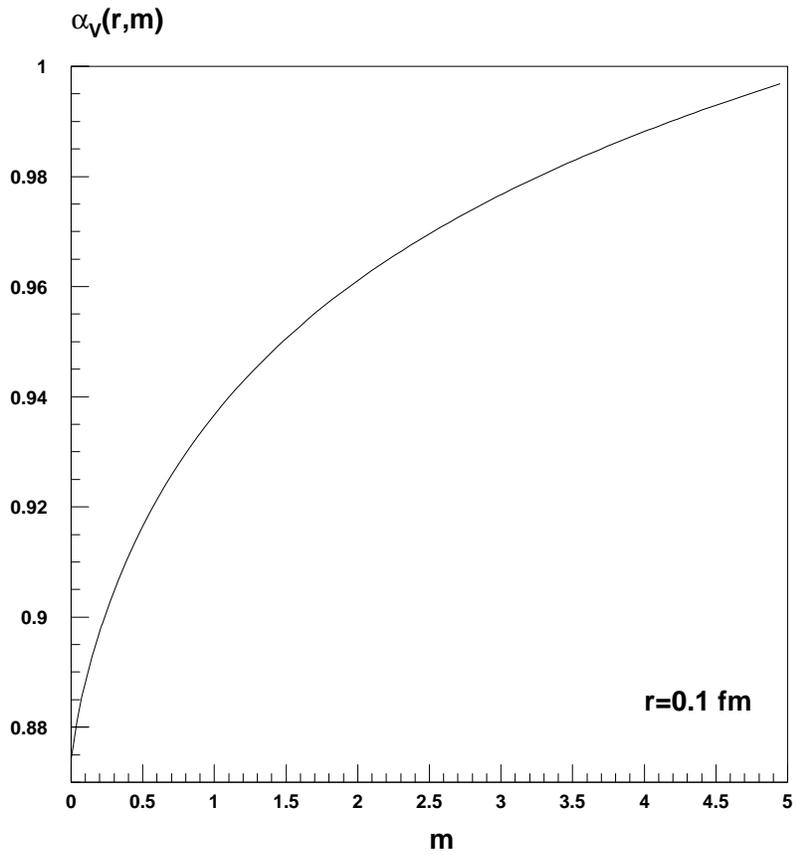,height=5.0in}
 \caption{The dependence of $\alpha_V(r,m)$ on the $\overline{\mbox{MS}}$
 mass parameter $m(\mu)$. The figure keeps $r=0.1$ fm fixed.
 The diagram is
 displayed for the ``natural choice'' of the renormalization scale $\mu=
 \frac{1}{r e^{\gamma_E}}$.}
 \label{fig:aVmm}
 \end{figure}
 The mass dependence of $\alpha_V(r,m)$ is displayed in Fig. \ref{fig:aVmm}.
 We keep $r$ fixed at $0.1$ fm and vary the 
 $\overline{\mbox{MS}}$-mass parameter between $0$ and $5$GeV. While
 perturbation theory might not be applicable for distances larger than
 $0.1$fm, the
 effect is significant and can amount to a few percent
 for $\overline{\mbox{MS}}$-masses up to $1$GeV for the physical charge
 $\alpha_V(r,m)$
 compared to the massless result. For the bottom quark mass the effect is
 of order 10 \% at $r=0.1$fm.

\section{The force between two static sources} \label{sec:af}

In this section we discuss the concept of defining the strong coupling
from the force between two static color-singlet sources \cite{gru}. In 
coordinate space the force is simply given by:
\begin{equation}
F(r,m) = - \frac{\partial V(r,m)}{\partial r} \equiv - C_F \frac{\alpha_F(r,m)}{r^2}\label{eq:aFdef}
\end{equation}
For large distances there is no required sign change for $\alpha_F(r,m)$ and
its accompanying $\beta$-function is unique. For the massless coupling
we can simply write:
\begin{equation}
\alpha_F(r)= \alpha_V(r) \left(1-2 \beta_V(\alpha_V(r)) \right) \label{eq:aFaV}
\end{equation}
where $\beta_V$ denotes the massless $\beta$-function in the V-scheme.
From Eq. \ref{eq:aFaV} it follows directly that the massless relation between
the $\alpha_F$-charge and the $\overline{\mbox{MS}}$-coupling to two loops
is given by
\begin{equation}
\alpha_F(r)=
\alpha_{\overline{\mbox{\tiny MS}}} (r^\prime) \left( 1 + f_1 
\frac{\alpha_{\overline{\mbox{\tiny MS}}} (r^\prime)}{\pi} +f_2
\left( \frac{\alpha_{\overline{\mbox{\tiny MS}}} (r^\prime)}{\pi} \right)^2
\right) \label{eq:aF0}
\end{equation}
where $r^\prime=r \exp (\gamma_E)$ and the constants are given by
\begin{eqnarray}
f_1&=&-\frac{35}{36} C_A + \frac{1}{9} T_F n_f \label{eq:f10} \\
f_2&=& \frac{1}{16} \left[ \left(- \frac{7513}{162}+\frac{229}{27} \pi^2 - \frac{1}{
4} \pi^4 + \frac{22}{3} \zeta_3 \right) C_A^2 + \left( \frac{3410}{81}-
\frac{88}{27} \pi^2-\frac{56}{3} \zeta_3 \right) C_AT_Fn_f \right. \nonumber \\
&& \left. - \left( \frac{31}{3}-16 \zeta_3 \right) C_FT_Fn_f - \left(
\frac{560}{81}-\frac{16}{27} \pi^2 \right) \left( T_Fn_f \right)^2 \right]
\label{eq:f20}
\end{eqnarray}
In general, the relation for the massive case is given by
\begin{eqnarray}
\alpha_F(r,m)&=&-r^2 \frac{\partial (\alpha_V(r,m)/r)}{\partial r} \nonumber \\
&=& 
\alpha_{\overline{\mbox{\tiny MS}}} (\mu) \left( 1 +
f_1(r,m,\mu) \frac{\alpha_{\overline{\mbox{\tiny MS}}}(\mu)}{\pi}
+f_2(r,m,\mu) \left( \frac{
\alpha_{\overline{\mbox{\tiny MS}}}(\mu)}{\pi}
\right)^2 \right) \label{eq:aFm}
\end{eqnarray}
at the two loop level.
Explicit expressions for $f_1(r,m,\mu)$ and $f_2(r,m,\mu)$ are given in the
appendix.
 \begin{figure}
 \center
 \epsfig{file=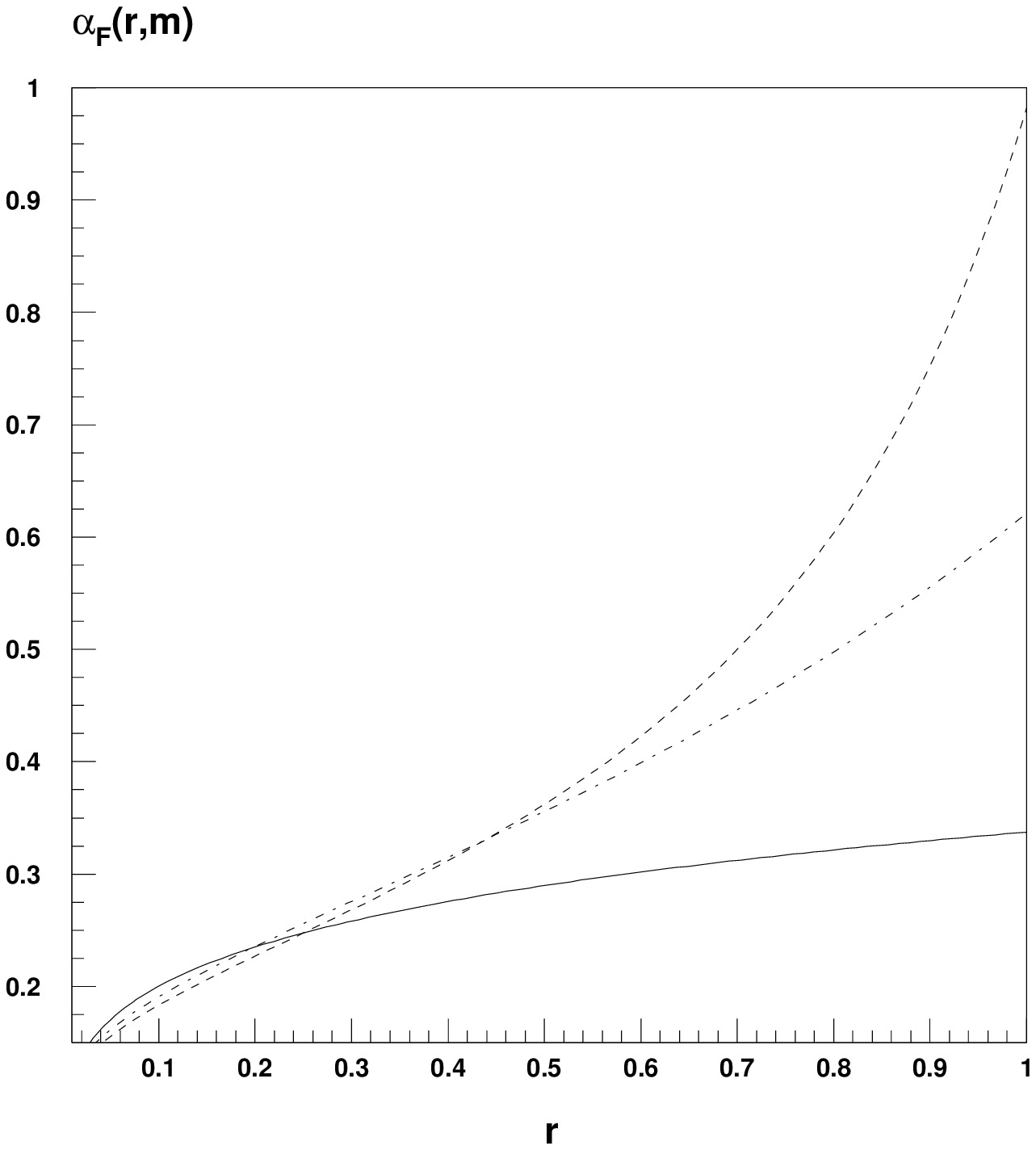,height=3.5in}
 \epsfig{file=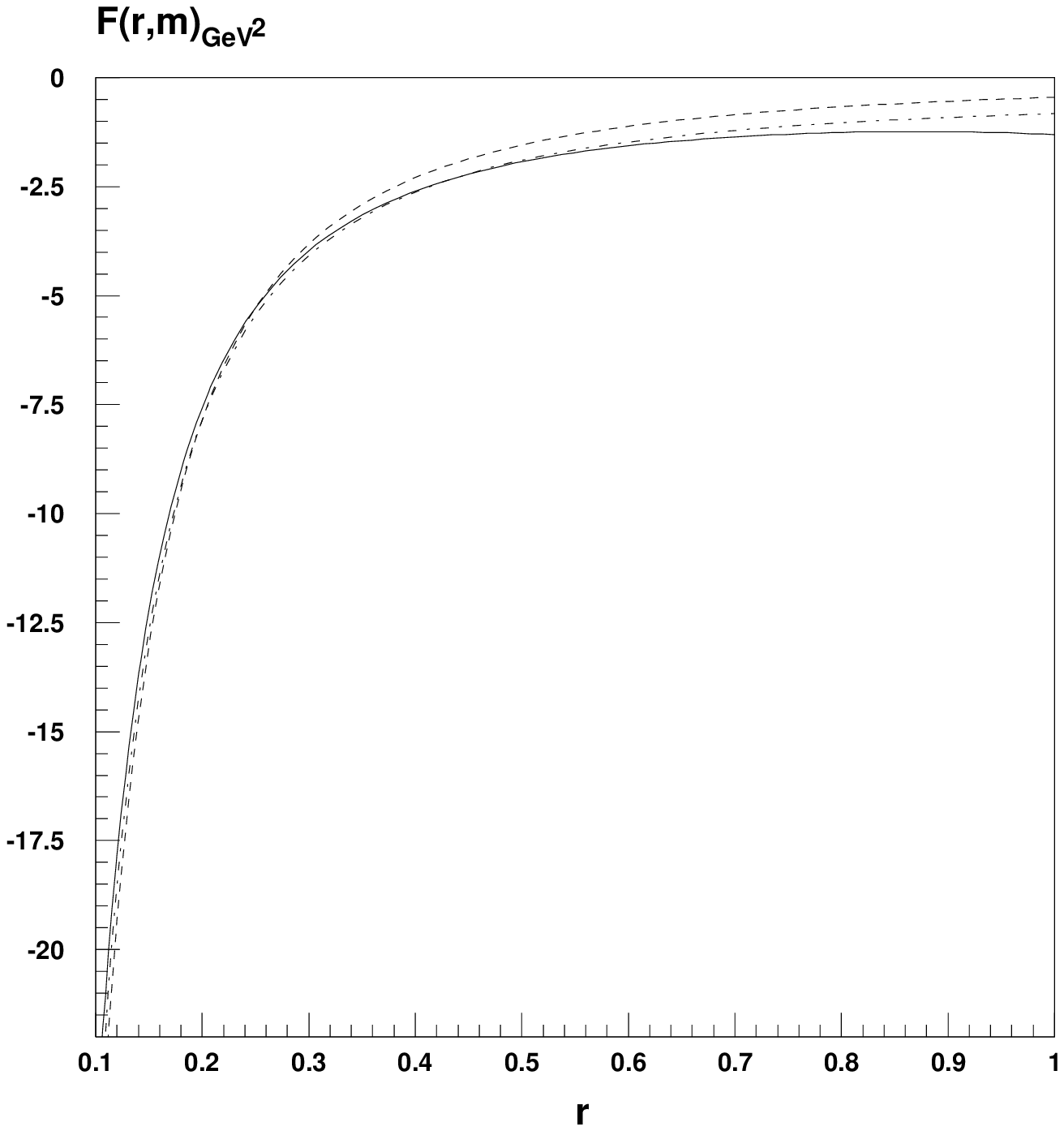,height=3.5in}
 \caption{The position dependence of $\alpha_F(r,m)$  and the force
 $F$ for different
 choices of the renormalization scale $\mu$. The scale choice corresponding
 to the respective lines is the same as in Fig. \ref{fig:aVmu}.
 The force in the lower plot is in GeV$^2$, $r$ in GeV$^{-1}$.}
 \label{fig:aFmu}
 \end{figure}
The distance dependence of the perturbative coupling definition of
$\alpha_F(r,m)$ is presented in Fig. \ref{fig:aFmu} for the natural
scale choice (dashed) and fixed $\mu=M_Z$ (solid). The former leads to a lower
value of $\alpha_F(r,m)$ at larger distances. Overall, it can be seen
that the coupling defined from the force is much smaller at a given
source-separation $r$ than $\alpha_V(r,m)$ (and even smaller than 
$\alpha_{\overline{\mbox{\tiny MS}}}(\mu=\frac{1}{re^{\gamma_E}})$).
Even at $r=0.2$ fm ($\equiv 1$GeV$^{-1}$) still seems to be 
acceptable from a perturbative point of 
view. This could make a comparison with lattice results very interesting.
Overall, the renormalization scale dependence is also strong for
$\alpha_F(r,m)$.
The lower graph displays the force itself in units of [GeV$^2$].
 \begin{figure}
 \center
 \epsfig{file=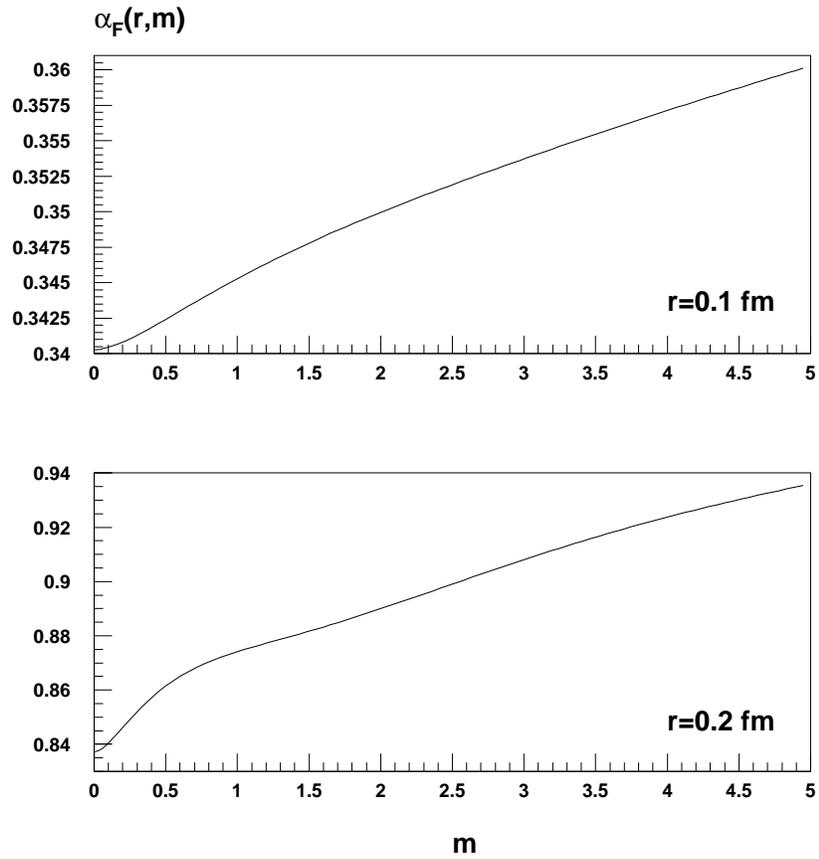,height=5.0in}
 \caption{The dependence of $\alpha_F(r,m)$ on the $\overline{\mbox{MS}}$
 mass parameter $\overline{m}$. The figure on the top keeps $r=0.1$ fm
 fixed while the one on the bottom has $r=0.2$ fm. Both diagrams are
 displayed to the ``natural choice'' of the renormalization scale $\mu=
 \frac{1}{r e^{\gamma_E}}$. The effect of mass terms is clearly 
 visible.}
 \label{fig:aFmm}
 \end{figure}
The mass dependence of $\alpha_F(r,m)$ is given in Fig. \ref{fig:aFmm}
analogously to Fig. \ref{fig:aVmm} by varying the $\overline{\mbox{MS}}$-mass
parameter between $0$ and $5$GeV and for two fixed values
of $r$. The dependence on the mass terms is not as large for small 
$\overline{\mbox{MS}}$
in comparison to $\alpha_V(r,m)$, however, can still be a few percent.
It can also be seen that
the effect for the  
$\overline{\mbox{MS}}$ bottom quark mass is roughly 7\% 
compared to the massless result at $r=0.1$fm, while it is about 10\%
at $r=0.2$fm.

\section{Conclusions} \label{sec:con}

We have calculated the coordinate space static QCD-potential through two
loops with quark masses. The result is obtained by reconstructing the 
exact momentum space Monte Carlo results from Ref. \cite{mel} by analytically 
fitting the Gell-Mann Low function. The reconstructed results are in
good (few percent) 
agreement with the fermionic results of the exact calculation.  
Based on these solutions
we have obtained the Fourier transform in coordinate space and the
force between the heavy quarks. The 
$\alpha_V(r,m)$ coupling itself grows quicker than $\alpha_{
\overline{\mbox{\tiny MS}}}(\mu)$, which is mainly due to its larger value
at $Q=M_Z$. 
The mass-dependence is strongest for small $\overline{\mbox{MS}}$ masses
up to 1GeV at large distances (0.1fm), where the perturbative regime breaks down.
At these distances the effect of keeping a 
massive charm or bottom quark can be a significant
effect, although perturbation theory might not be trustworthy 
anymore at distances
of order 0.1 fm. An important indicator of this fact is the strong 
renormalization scale dependence from uncanceled higher order terms which
can lead to substantially different results.

For the corresponding
definition of $\alpha_F(r,m)$ we find also a strong dependence on the mass
terms. The effect is not as large for small $\overline{\mbox{MS}}$ masses,
however can still be a few percent. At 0.2fm the effect is of order 10\%
for the bottom quark mass.
The renormalization scale dependence from uncanceled higher order terms
is also large for $\alpha_F(r,m)$.
The overall smallness of this coupling
parameter, however, indicates that it could serve as a suitable quantity
to compare the perturbative and non-perturbative approaches even above
scales of ${\cal O}(0.1$ fm).

In terms of practical applications, the relatively simple form of Eqs.
\ref{eq:aVm} and \ref{eq:aVrm} makes it possible to investigate the
uncertainty of a massive charm contribution in the determination of the
bottom quark mass \cite{hmel}.

\section*{Acknowledgments}
The author would like to acknowledge interesting discussions with R.~Sommer and 
A.H.~Hoang and to thank A.~Denner for carefully reading the manuscript.

\section*{Appendix} \label{sec:app}

\subsection*{Explicit two loop results with quark masses for $\alpha_F(r,m)$}

In this appendix we give the explicit expressions for the two loop results
for the strong coupling parameter defined from the force between two static
color singlet sources. The general definition is given in Eq. \ref{eq:aFm} and
is given by:
\begin{eqnarray}
&&f_1(r,m,\mu)= C_A \left( -\frac{35}{36}+ \frac{11}{6} ( \log (\mu \; r) +\gamma_E
) \right) + T_F \left[ - \frac{5}{9} + \frac{2}{3} \exp \left( - \sqrt{a_0} \;
m \; r \right) \right. \nonumber \\ && + 
\left. \frac{1}{3} \left( \log \frac{a_0 m^2}{
\mu^2} + 2 \; \mbox{Ei} \left( 1, \sqrt{a_0} \; m \; r \right) \right) \right] \\
&&f_2(r,m,\mu)= \frac{C_A^2}{16} \left( - \frac{7513}{162} + \frac{229}{27} \pi^2-\frac{\pi^4}{4}
+ \frac{22}{3} \zeta_3 - \frac{928}{27} \left(  
\log ( \mu \; r ) +\gamma_E \right) \right. \nonumber \\ && \left. 
+ \frac{484}{9} \left( \log ( \mu \;r )+\gamma_E \right)^2 \right) \nonumber \\ && - 
T_F \left[
\frac{19}{6} \left\{-\frac{1}{2} \left(
\exp \left(
-\sqrt{ \frac{a_3+\sqrt{a_3^2-4 a_2}}{2 a_2}} \;m \;r \right) +
\exp \left( -\sqrt{ \frac{a_3-\sqrt{a_3^2-4 a_2}}{2 a_2}} \;m \;r \right) 
\right) \right. \right. \nonumber \\
&& + \left. \frac{ \frac{a_3}{2}-a_1}{\sqrt{a_3^2-4a_2}} \left( \exp
\left( - \sqrt{a_3-\sqrt{a_3^2-4a_2}} \;m\;r \right) - \exp 
\left( - \sqrt{a_3+\sqrt{a_3^2-4a_2}} \;m\;r \right) \right) \right\}
\nonumber \\ && - \frac{55}{27} C_A - \left( \frac{31}{27} C_A-\frac{20}{27}
T_F \right) \exp \left( - \sqrt{a_0} \; m \;r \right) 
- \frac{22}{9} C_A \left( \log ( \mu \;r) +\gamma_E \right) 
- \frac{11}{9} \frac{C_A}{\pi} \;r\; {\cal I}_3 \nonumber \\ &&
+ \frac{2}{9} \frac{T_F}{\pi} \;r\; {\cal I}_4  
- \left\{ -\frac{C_A}{16} \left( \frac{1798}{81}+\frac{56}{3}
\zeta_3 \right) - \frac{C_F}{16} \left( \frac{55}{3}-16 \zeta_3 \right)
+ \frac{25}{81} T_F \right. \nonumber \\ &&
+ \frac{19}{6} \left\{ \frac{1}{4} \left( \log \frac{(a_3+\sqrt{a_3^2-4a_2})m^2}{
2a_2 \mu^2} + 2 \mbox{Ei} \left( 1, \sqrt{ \frac{a_3+ \sqrt{a_3^2-4a_2}}{2a_2}} \; m \;
r \right) \right. \right. \nonumber \\ &&
+ \left. \log \frac{(a_3-\sqrt{a_3^2-4a_2})m^2}{
2a_2 \mu^2} + 2 \mbox{Ei} \left( 1, \sqrt{ \frac{a_3- \sqrt{a_3^2-4a_2}}{2a_2}} \; m \;
r \right) \right) \nonumber \\ && + \frac{\frac{a_3}{4}-\frac{a_1}{2}}{\sqrt{a_3^2-4a_2}} 
\left( \log \frac{a_3+ \sqrt{a_3^2-4a_2}}{a_3- \sqrt{a_3^2-4a_2}}
+2 \mbox{Ei} \left( 1, \sqrt{a_3+ \sqrt{a_3^2-4a_2}} \; m \; r \right) \right. 
\nonumber \\ && \left. \left. - 2 \mbox{Ei} \left( 1, \sqrt{a_3+ \sqrt{a_3^2-4a_2}} 
\; m \; r \right) \right) \right\} - \frac{55}{27} C_A ( \log ( \mu \;r) +
\gamma_E ) \nonumber \\ &&
+ \left( \frac{31}{54}C_A-\frac{10}{27}T_F \right) \left( \log \frac{a_0 m^2}{
\mu^2} + 2 \mbox{Ei} \left( 1, \sqrt{a_0} \; m \; r \right) \right) 
\nonumber \\ && - \frac{11}{36}
C_A \left( 4 \left( \log ( \mu r ) + \gamma_E \right)^2 + \frac{\pi^2}{3} \right)
\left. \left. - \frac{11}{9} \frac{C_A}{\pi} {\cal I}_1 + \frac{2}{9} 
\frac{T_F}{\pi} {\cal I}_2 \right\} \right]
\end{eqnarray}
where
\begin{eqnarray}
{\cal I}_3 (r,m,\mu) &\equiv& \int^\infty_0 \log \frac{ Q^2+a_0m^2}{Q\mu} \log \frac{Q^2}{
\mu^2} \cos ( Q \; r) dQ \label{eq:ir3} \\
{\cal I}_4 (r,m,\mu) &\equiv& \int^\infty_0 \log^2 \frac{ Q^2+a_0m^2}{\mu^2}
\cos ( Q \; r) dQ \label{eq:ir4}
\end{eqnarray}
For the running $\overline{\mbox{MS}}$-mass one needs to add the following
contributions:
\begin{eqnarray}
f_2(r,m(\mu),\mu)&=& f_2(r,m,\mu)- \frac{1}{6} C_F T_F \left( 4 + 3 \log
\frac{\mu^2}{m^2} \right) \left( \sqrt{a_0} m \; r\;  \exp \left( - \sqrt{a_0}
\; m \; r \right) \right. \nonumber \\ &&
\left. - \left( 1 - \exp \left( - \sqrt{ a_0} \; m \; r \right) \right) 
\right)
\end{eqnarray}

\subsection*{Mixed massive and massless fermionic
contributions at the two loop level}

\begin{center}
\begin{figure}
\centering
\epsfig{file=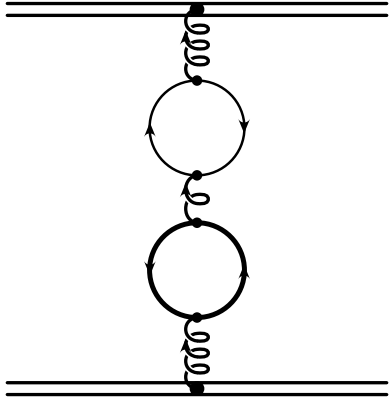,width=2.0in}
\caption{The Feynman diagram which can contain both massless and massive (thick
line)
quark loops or two flavors of different non-zero mass at the two loop level.
} \label{fig:tfvp}
\end{figure}
\end{center}
In this part of the appendix we briefly comment on the type of contribution
depicted in Fig. \ref{fig:tfvp}. This diagram can contain the same flavor
in both loops as well as different ones. In the former case, the contribution
is already contained in the discussion above. In section \ref{sec:rc} we
demonstrated that for the Gell-Mann Low function only one flavor needs to
be considered. On the level of the potential, however, we have also the
latter case represented in Fig. \ref{fig:tfvp}. 

In a practical situation, we can often set the masses of the three lightest
quarks, u,d and s to zero. In this case the Feynman graph \ref{fig:tfvp}
contributes with a multiplicity factor two and we find the resulting
expression:
\begin{eqnarray}
2\times \mbox{Fig.13}_{0,m} \!\!\!\!&=&\!\!\!\! -2 \frac{4 \pi C_F}{Q^2} \left( \frac{31}{36} C_A - \frac{5}{9} T_F n_f+
\frac{1}{4} \beta_0 \log \frac{\mu^2}{Q^2} \right) \times \nonumber \\
\!\!\!\!&&\!\!\!\! \left( \frac{31}{36} C_A - \frac{5}{9} T_F + \frac{11}{12} C_A \log
\frac{\mu^2}{Q^2} - \frac{1}{3} T_F \log \frac{\mu^2}{Q^2+a_0m^2} \right)
\frac{\alpha_{\overline{\tiny{\mbox{MS}}}}^3(\mu)}{\pi^2}
\end{eqnarray}
In the case of two non-zero and different quark loops we find analogously:
\begin{eqnarray}
2\times \mbox{Fig.13}_{m,m^\prime} \!\!\!\!&=&\!\!\!\! -2 \frac{4 \pi C_F}{Q^2} \left(  
\frac{31}{36} C_A - \frac{5}{9} T_F + \frac{11}{12} C_A \log
\frac{\mu^2}{Q^2} - \frac{1}{3} T_F \log \frac{\mu^2}{Q^2+a_0m^2} \right)
\times \nonumber \\
\!\!\!\!&&\!\!\!\! \left( \frac{31}{36} C_A - \frac{5}{9} T_F + \frac{11}{12} C_A \log
\frac{\mu^2}{Q^2} - \frac{1}{3} T_F \log \frac{\mu^2}{Q^2+a_0{m^\prime}^2} 
\right)
\frac{\alpha_{\overline{\tiny{\mbox{MS}}}}^3(\mu)}{\pi^2}
\end{eqnarray}
In the above expressions we used again the approximate one loop solution for
the vacuum approximation. If desired, it can of course be substituted with
the exact function containing more complicated expressions \cite{bgmr}.
We also already divided by a factor $i$ to obtain the contribution to the
potential.

\end{document}